\documentclass[
 reprint,
 amsmath,
 amssymb,
 aps,
 superscriptaddress
]{revtex4-2}

\usepackage{array}
\usepackage{tabularx}
\usepackage{makecell}
\usepackage{xcolor}
\usepackage{graphicx}
\usepackage{dcolumn}
\usepackage{bm}
\usepackage{float}
\usepackage{soul}
\usepackage{multirow}

\usepackage[
    colorlinks=true,
    citecolor=blue,
    linkcolor=blue,
    urlcolor=blue
]{hyperref}

\begin{document}

\preprint{APS/123-QED}

\title{Spin-Group versus Magnetic-Group Selection Rules for Nonlinear Photocurrents in Compensated Ferrimagnets}

\author{Fu Li}
\email{fu.li@tu-darmstadt.de}
\affiliation{
Institute of Materials Science, Technical University of Darmstadt,
64287 Darmstadt, Germany
}

\author{Vikrant Chaudhary}
\affiliation{
Institute of Materials Science, Technical University of Darmstadt,
64287 Darmstadt, Germany
}

\author{Weiyi Pan}
\affiliation{Institute for Theoretical Physics, University of Regensburg, 93040 Regensburg, Germany}

\author{Hongbin Zhang}
\email{hongbin.zhang@tu-darmstadt.de}
\affiliation{
Institute of Materials Science, Technical University of Darmstadt,
64287 Darmstadt, Germany
}

\author{Harish K. Singh}
\email{harish.singh@plus.ac.at}
\affiliation{Department of Chemistry and Physics of Materials,
University of Salzburg, Jakob-Haringer-Strasse 2a,
5020 Salzburg, Austria}

\begin{abstract}
Optical excitation offers a contact-free, ultrafast route to charge and spin currents, with symmetry determining which nonlinear photocurrents are allowed and whether they carry charge, spin, or both. Spin-group symmetry describes the limit without spin--orbit coupling (SOC), whereas magnetic-group symmetry accounts for SOC. We show that the two descriptions yield different selection rules for both charge and spin photocurrents in the multiferroic compensated ferrimagnet GaFeO$_3$. Without SOC, only linear shift and circular injection currents are allowed, with identical components for charge and spin, so every charge current is spin polarized along the moments. SOC leaves these charge currents nearly unchanged, activates magnetic linear injection and circular shift charge currents, and enables pure spin currents polarized perpendicular to the moments. Ferroelectric switching reverses the linear shift and circular injection charge currents and their spin polarization. These findings establish the importance of magnetic-group symmetry for a complete description of nonlinear charge and spin photocurrents in compensated ferrimagnets.
\end{abstract}

\maketitle

\underline{\it Introduction--}
Generating spin currents is a central goal of spintronics~\cite{vzutic2004spintronics,fert2008nobel,maekawa2023spin}. The spin Hall effect, spin pumping, and the spin Seebeck effect convert charge currents, magnetization dynamics, and thermal gradients into spin currents, respectively~\cite{sinova2015spin,li2026high,tserkovnyak2005nonlocal,uchida2008observation}. Optical excitation can instead provide a contact-free, ultrafast route to spin-current generation~\cite{hubner2003direct,xu2021pure}. The bulk photovoltaic effect (BPVE) generates a dc charge photocurrent through a second-order optical response under uniform illumination when inversion symmetry is broken by the crystal structure or magnetic order, and the bulk spin photovoltaic effect extends it to spin transport~\cite{young2013prediction,xu2021pure,yang2026symmetry,li2026magnetic}. Microscopically, the second-order photocurrent is governed by two distinct mechanisms. Shift currents (SC) reflect real-space electronic displacements during interband transitions, whereas injection currents (IC) originate from an asymmetric distribution of photoexcited carriers in velocity space~\cite{sipe2000second,panday2019injection}. Both mechanisms can be driven by linearly or circularly polarized light, yielding linear and circular shift currents (LSC and CSC) and linear and circular injection currents (LIC and CIC).

For charge currents, the four channels are classified as even or odd under time reversal $\mathcal{T}$. LSC and CIC are $\mathcal{T}$-even and occur in any noncentrosymmetric crystal, whereas LIC and CSC are $\mathcal{T}$-odd magnetic responses that require broken $\mathcal{T}$~\cite{zhang2019switchable,watanabe2021chiral}. In $\mathcal{PT}$-symmetric antiferromagnets, where $\mathcal{T}$ is broken but $\mathcal{PT}$ is preserved, only this magnetic pair survives~\cite{zhang2019switchable,watanabe2021chiral}, whereas a polar magnet without $\mathcal{PT}$ can host all four~\cite{li2026magnetic}. Which of them appear depends on whether SOC is included.

Whether these channels carry pure spin, spin-polarized charge, or pure charge currents depends on the symmetry. Without SOC, spin point groups (SPGs) describe independent spin and lattice rotations, and spin along the collinear ordering direction $\hat{\mathbf{n}}$ is conserved~\cite{litvin1974spin,liu2022spin}. Charge and longitudinal spin responses are the sum and difference of the spin-channel responses, with effective time reversal within each channel allowing only LSC and CIC~\cite{dong2025crystal}. SOC couples spin to the lattice, requiring magnetic point groups (MPGs), under which spin transforms as a $\mathcal{T}$-odd axial vector~\cite{birss1964symmetry,gallego2019automatic}. This can activate LIC and CSC~\cite{zhang2019switchable,watanabe2021chiral} and enable pure spin currents where mirrors forbid charge flow~\cite{xu2021pure}.


Without SOC, a symmetry-protected pure spin photocurrent requires an operation interchanging the two spin channels, such as $\mathcal{PT}$ in antiferromagnets~\cite{fei2021pt} or a rotation or mirror in altermagnets, where only tensor components odd under that spatial operation support pure spin currents~\cite{young2013prediction,smejkal2022beyond,dong2025crystal,yang2025giant}. SOC can spoil this purity in antiferromagnets by admitting charge LIC and CSC~\cite{zhang2019switchable,watanabe2021chiral}, whereas altermagnets can retain it~\cite{dong2025crystal}. Ferromagnets need no such symmetry and carry spin-polarized charge photocurrents~\cite{liu2023generation}, but their magnetization produces stray fields~\cite{baltz2018antiferromagnetic}. Compensated ferrimagnets (CFIMs) retain this ferromagnet-like response at zero net magnetization~\cite{yuan2024nonrelativistic,liu2025two,chaudhary2026coexistence}. Their antiparallel sublattices are crystallographically inequivalent, so no symmetry interchanges the spin channels, and compensation is instead enforced by electron filling~\cite{guo2025luttinger}.

\begin{figure}[!htp]
\centering
\includegraphics[width=1.0\linewidth]{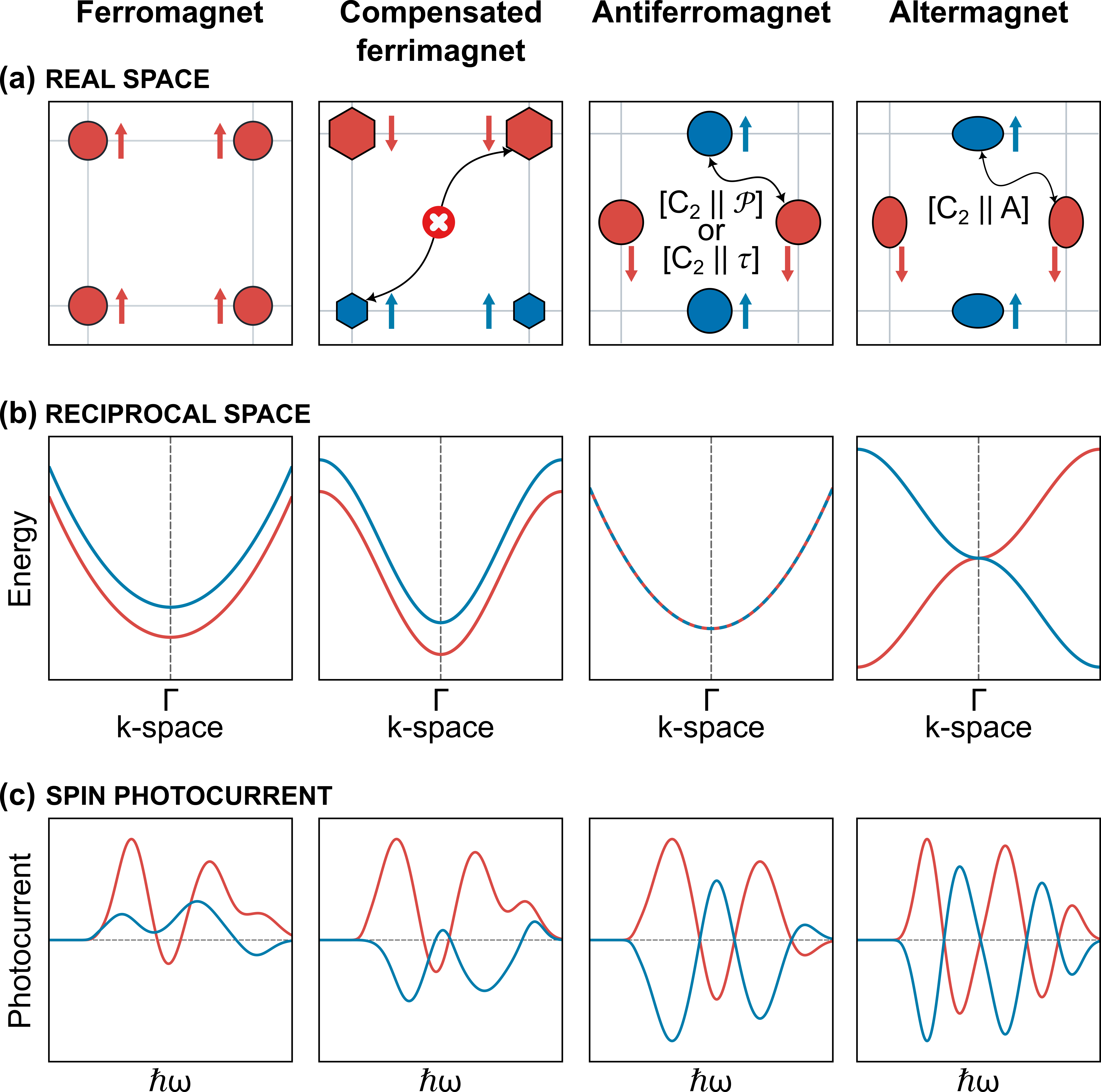}
\caption{Schematic comparison without SOC.
(a) Opposite-spin sublattices are unrelated by symmetry in CFIMs, linked by inversion $\mathcal{P}$ or translation $\tau$ in AFMs, and by a rotation or mirror $A$ in ALMs. Here, $C_2$ denotes a $180^\circ$ spin rotation.
(b) Spin-split FM and CFIM bands, spin-degenerate AFM bands, and alternating ALM spin splitting.
(c) Spin-channel photocurrents are independent in FMs and CFIMs but equal and opposite in $\mathcal{P}$-linked AFMs and ALM components, yielding pure spin currents ($\sigma_{\uparrow}+\sigma_{\downarrow}=0$).}
\label{fig:schematic}
\end{figure}

However, SPG operations in CFIMs never interchange the two spin channels, so spin and charge currents share the same allowed components. The nonrelativistic response is therefore generically a spin-polarized charge photocurrent, with no symmetry-protected pure spin channel. The role of SOC is thereby reversed, as only SOC enables symmetry-protected pure spin photocurrents and only the MPG identifies their allowed components. For charge currents, SOC is likewise required for the magnetic LIC and CSC. Polar CFIMs also allow ferroelectric control, because second-order charge photoconductivities, unlike the spin Hall conductivity, are odd under inversion. In two-dimensional ferroelectric altermagnets~\cite{yang2025giant,yang2026symmetry}, such switchable photocurrents remain in-plane; in a bulk film they can flow along its normal, as needed for vertical spin injection.

\begin{figure}[!htp]
\centering
\includegraphics[width=0.98\linewidth]{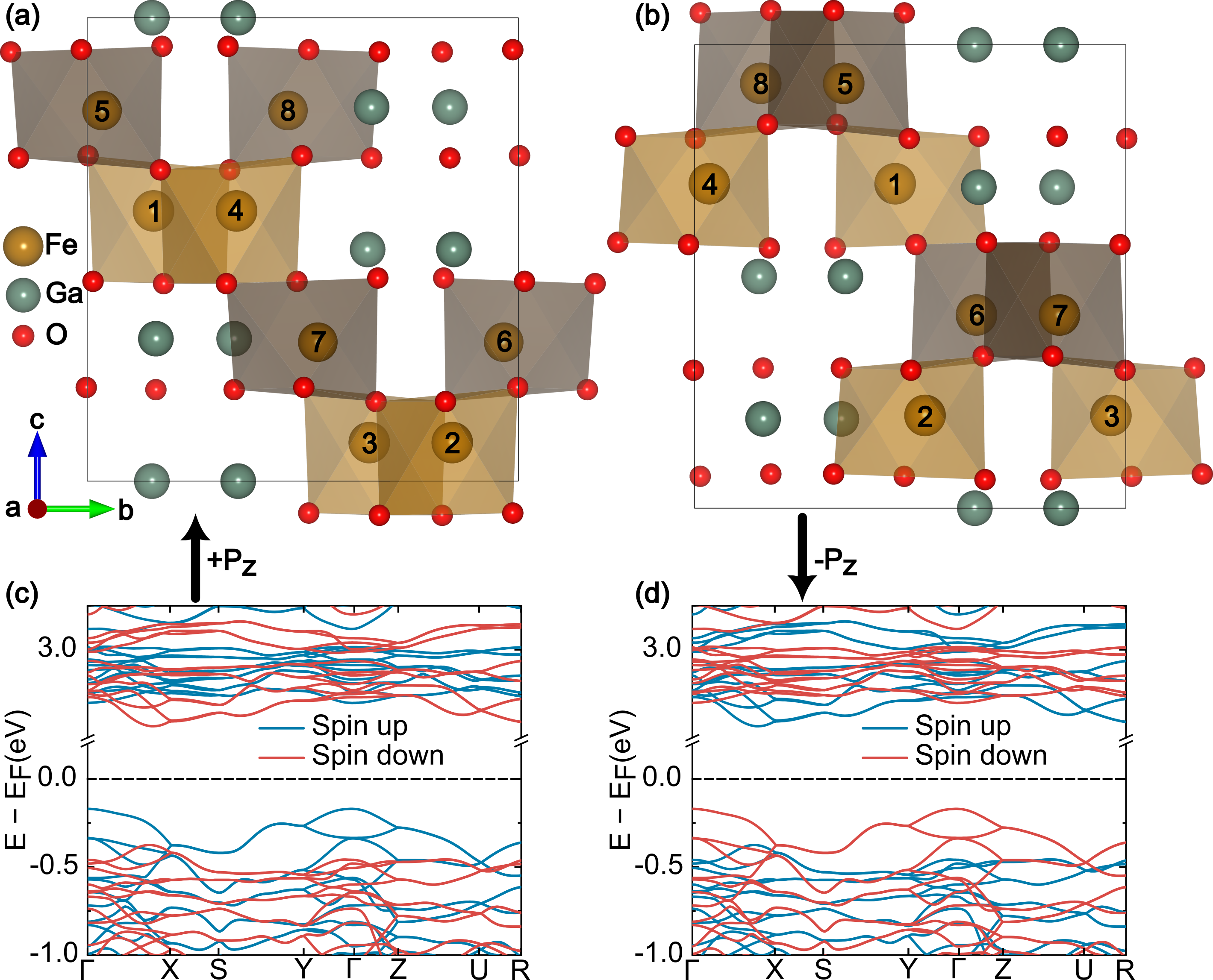}
\caption{GaFeO$_3$: (a) crystal and magnetic structure; (b) spin-resolved bands. Electronic structures are calculated without spin--orbit coupling, with the Fermi energy set to zero. Light golden and dark octahedra mark Fe sites with spin-up and spin-down moments, respectively.
}
\label{fig:GaFeO3}
\end{figure}

Here, we derive the SPG and MPG selection rules for all four charge and spin photocurrent channels in the polar CFIM GaFeO$_3$. From first principles, we calculate the charge photocurrents with and without SOC and the spin photocurrents without SOC. Spin photocurrents with SOC are analyzed by symmetry alone. Without SOC, the spin LSC and CIC are polarized along the Fe moments and comparable to the charge currents. SOC leaves the charge LSC and CIC nearly unchanged and activates the charge LIC and CSC. The MPG predicts spin LIC and CSC and pure spin currents polarized perpendicular to the moments. Ferroelectric reversal, which leaves the Fe moments fixed, reverses the charge LSC and CIC and their spin polarization. Our results distinguish dominant exchange-driven spin-polarized currents from SOC-enabled magnetic charge and pure spin currents and establish the need for both symmetry descriptions.

\underline{\it Methods--} First-principles calculations were performed using the Vienna \textit{Ab initio} Simulation Package (VASP)~\cite{kresse1993ab,kresse1994ab,kresse1996efficiency} with the Perdew--Burke--Ernzerhof exchange--correlation functional~\cite{perdew1996generalized}. Maximally localized Wannier functions were constructed using Wannier90~\cite{mostofi2014updated} to obtain a Wannier-based tight-binding Hamiltonian (Figs.~\textcolor{blue}{S1} to \textcolor{blue}{S3}). Further computational details are given in Section~\textcolor{blue}{S1} of the Supporting Information (SI).

A dc photocurrent arises at second order in the optical field. Within the electric-dipole approximation, the shift and steady-state injection currents are
\begin{equation}
J_i^{\mathrm{shift}}
=
\sigma_{ijk}(\omega)E_j(\omega)E_k(-\omega),
\end{equation}
\begin{equation}
J_i^{\mathrm{injection}}
=
\tau\,\eta_{ijk}(\omega)E_j(\omega)E_k(-\omega),
\end{equation}
where $E_j(\omega)$ and $E_k(-\omega)=E_k^{*}(\omega)$ are optical-field components, $i$ is the current direction, and $\tau$ is the relaxation time. The response tensors and the definitions of charge and spin photocurrents are also given in Section~\textcolor{blue}{S1}.
\begin{table*}[htbp]
\centering
\caption{Symmetry-allowed tensor components for second-order charge and spin photocurrents in GaFeO$_3$. The SPG and MPG $mm'2'$ constrain responses without and with SOC, respectively. In $\sigma_{i;jk}$ ($\eta_{i;jk}$) and $\sigma^s_{i;jk}$ ($\eta^s_{i;jk}$), $i$ is the current direction, $j,k$ are the optical-field directions, and $s$ is the spin-polarization direction. In the Jahn symbols, $V$ and $M$ denote polar-vector and independent spin-vector indices, respectively, while $e$ marks axial character and $a$ supplies an additional time-reversal-odd factor. Square (curly) brackets denote symmetric (antisymmetric) optical-index pairs. Only one representative of each optical-index pair is listed: $T_{i;jk}=T_{i;kj}$ for linear channels and $T_{i;jk}=-T_{i;kj}$ for circular channels, where $T=\sigma,\eta$ (likewise for $T^s$). Components marked $0$ are symmetry forbidden. All components follow from the symmetry analysis in Sec.~\textcolor{blue}{S2} and Table~\textcolor{blue}{S1}, with their transformations listed in Tables~\textcolor{blue}{S2--S17}.}
\label{tab:gafeo3_photocurrent}

\small
\renewcommand{\arraystretch}{1.2}
\setlength{\tabcolsep}{3pt}
\newlength{\compgap}\setlength{\compgap}{10pt}
\newcommand{\cellstack}[1]{%
  \begin{tabular}[c]{@{}c@{}}\rule{0pt}{3ex}#1\rule[-1.5ex]{0pt}{0pt}\end{tabular}}
\newcommand{\chan}[2]{\begin{tabular}[c]{@{}l@{}}#1\\#2\end{tabular}}

\begin{tabular*}{\textwidth}{@{}l cc
  cc @{\hspace{\compgap}\extracolsep{\fill}}c@{\hspace{\compgap}}c @{\extracolsep{0pt}\hspace{2\tabcolsep}}
  cc @{\hspace{\compgap}\extracolsep{\fill}}c@{\extracolsep{0pt}}}
\hline\hline
\multirow{3}{*}{Channel}
  & \multicolumn{2}{c}{Spin}
  & \multicolumn{4}{c}{Spin}
  & \multicolumn{3}{c}{Charge} \\
  & \multicolumn{2}{c}{(without SOC)}
  & \multicolumn{4}{c}{(with SOC)}
  & \multicolumn{3}{c}{(without / with SOC)} \\
\cline{2-3}\cline{4-7}\cline{8-10}
  & Jahn & $s=x$
  & Jahn & $s=x$ & $s=y$ & $s=z$
  & Jahn & without & with \\
\hline

\chan{LSC}{}
  & $MV[V^2]$
  & \cellstack{$\sigma^x_{x;xz},\ \sigma^x_{y;yz}$\\
               $\sigma^x_{z;xx},\ \sigma^x_{z;yy}$\\
               $\sigma^x_{z;zz}$}
  & $aeVV[V^2]$
  & \cellstack{$\sigma^x_{x;xz},\ \sigma^x_{y;yz}$\\
               $\sigma^x_{z;xx},\ \sigma^x_{z;yy}$\\
               $\sigma^x_{z;zz}$}
  & \cellstack{$\sigma^y_{x;yz},\ \sigma^y_{y;xz}$\\
               $\sigma^y_{z;xy}$}
  & \cellstack{$\sigma^z_{x;xx},\ \sigma^z_{x;yy}$\\
               $\sigma^z_{x;zz},\ \sigma^z_{y;xy}$\\
               $\sigma^z_{z;xz}$}
  & $V[V^2]$
  & \cellstack{$\sigma_{x;xz},\ \sigma_{y;yz}$\\
               $\sigma_{z;xx},\ \sigma_{z;yy}$\\
               $\sigma_{z;zz}$}
  & \cellstack{$\sigma_{x;xz},\ \sigma_{y;yz}$\\
               $\sigma_{z;xx},\ \sigma_{z;yy}$\\
               $\sigma_{z;zz}$} \\
\hline

\chan{LIC}{}
  & $aMV[V^2]$
  & $0$
  & $eVV[V^2]$
  & \cellstack{$\eta^x_{x;xy},\ \eta^x_{z;yz}$\\
               $\eta^x_{y;xx},\ \eta^x_{y;yy}$\\
               $\eta^x_{y;zz}$}
  & \cellstack{$\eta^y_{x;xx},\ \eta^y_{x;yy}$\\
               $\eta^y_{x;zz},\ \eta^y_{y;xy}$\\
               $\eta^y_{z;xz}$}
  & \cellstack{$\eta^z_{x;yz},\ \eta^z_{y;xz}$\\
               $\eta^z_{z;xy}$}
  & $aV[V^2]$
  & $0$
  & \cellstack{$\eta_{x;xy},\ \eta_{z;yz}$\\
               $\eta_{y;xx},\ \eta_{y;yy}$\\
               $\eta_{y;zz}$} \\
\hline

\chan{CIC}{}
  & $MV\{V^2\}$
  & $\eta^x_{x;xz},\ \eta^x_{y;yz}$
  & $aeVV\{V^2\}$
  & $\eta^x_{x;xz},\ \eta^x_{y;yz}$
  & \cellstack{$\eta^y_{x;yz},\ \eta^y_{y;xz}$\\
               $\eta^y_{z;xy}$}
  & $\eta^z_{y;xy},\ \eta^z_{z;xz}$
  & $V\{V^2\}$
  & $\eta_{x;xz},\ \eta_{y;yz}$
  & $\eta_{x;xz},\ \eta_{y;yz}$ \\
\hline

\chan{CSC}{}
  & $aMV\{V^2\}$
  & $0$
  & $eVV\{V^2\}$
  & $\sigma^x_{x;xy},\ \sigma^x_{z;yz}$
  & $\sigma^y_{y;xy},\ \sigma^y_{z;xz}$
  & \cellstack{$\sigma^z_{x;yz},\ \sigma^z_{y;xz}$\\
               $\sigma^z_{z;xy}$}
  & $aV\{V^2\}$
  & $0$
  & $\sigma_{x;xy},\ \sigma_{z;yz}$ \\
\hline\hline
\end{tabular*}
\end{table*}
\underline{\it Symmetry Classification--}
Without SOC, spin-group operations are written as $[U\thinspace\Vert\thinspace R]$, where $U$ and $R$ act independently in spin and real space, respectively. As shown in Fig.~\ref{fig:schematic}(a), opposite-spin sublattices are connected by $[C_2\thinspace\Vert\thinspace\mathcal{P}]$ or $[C_2\thinspace\Vert\thinspace\tau]$ in AFMs and $[C_2\thinspace\Vert\thinspace A]$ in ALMs. Here, $C_2$ reverses the moments through a $180^\circ$ spin rotation, $\mathcal{P}$ denotes inversion, $\tau$ translation, and $A$ a crystal rotation or mirror. In CFIMs, the opposite-spin sublattices are crystallographically inequivalent and not connected by any symmetry. These real-space sublattice relations are reflected in the spin-resolved bands (Fig.~\ref{fig:schematic}(b)). AFMs have spin-degenerate bands, ALMs exhibit alternating spin splitting that vanishes at $\Gamma$, and ferromagnets (FMs) and CFIMs allow spin splitting throughout the Brillouin zone, including at $\Gamma$.

These symmetry relations also govern how the two spin channels contribute to the photocurrent (Fig.~\ref{fig:schematic}(c)). The sum and difference of the spin-channel contributions give the charge and spin responses, $\sigma_{\mathrm{charge}}=\sigma_{\uparrow}+\sigma_{\downarrow}$ and $\sigma_{\mathrm{spin}}=\sigma_{\uparrow}-\sigma_{\downarrow}$, respectively, with analogous definitions for $\eta$. The operation $[C_2\thinspace\Vert\thinspace\mathcal{P}]$ enforces equal and opposite channel currents, yielding a pure spin response, whereas $[C_2\thinspace\Vert\thinspace\tau]$ enforces identical currents and zero spin response. In ALMs, the spatial operation $A$ selects the tensor components supporting pure spin currents~\cite{dong2025crystal}. In FMs and CFIMs, no symmetry exchanges the spin channels, so their responses generally differ in magnitude and spectrum. A finite spin photocurrent can therefore arise even when the two spin channels carry unequal currents in the same direction. CFIMs thus combine FM-like optical spin selectivity with zero net magnetization, while the crystal symmetries determine the allowed components.

\underline{\it Crystal, Magnetic, and Electronic Structure--}
GaFeO$_3$ crystallizes in polar orthorhombic $Pna2_1$ (No.~33, point group $mm2$), with eight formula units per cell and polarization along the $c$ axis. Its two Ga, two Fe, and six O crystallographically inequivalent sites occupy distinct $4a$ Wyckoff orbits~\cite{arima2004structural}, and we consider the fully cation-ordered stoichiometric structure. Both Fe sites have sixfold oxygen coordination, forming distorted FeO$_6$ octahedra, and carry antiparallel local moments of approximately $+4.257\,\mu_{\mathrm{B}}$ and $-4.256\,\mu_{\mathrm{B}}$ per Fe atom along $a$ (Fig.~\textcolor{blue}{S4}). Including the induced Ga and O moments, the total cell moment vanishes without SOC and remains negligible with SOC. In the collinear limit without SOC, equal integer occupations of the two insulating spin channels give $M_{\mathrm{spin}}=\mu_{\mathrm{B}}(N_{\uparrow}-N_{\downarrow})=0$, consistent with Luttinger electron counting~\cite{guo2025luttinger}. The magnetic structure has $\mathbf{k}=(0,0,0)$ and magnetic space group $Pna'2_1'$ (BNS No.~33.147). Its magnetic point group is of type $m'm2'$, oriented as $mm'2'$ in the present axes ($x\parallel a$, $y\parallel b$, $z\parallel c$), with operations $\{E,m_x,m_y',2_z'\}$.

Ferroelectric switching has been demonstrated in GaFeO$_3$ thin films~\cite{mukherjee2013room,song2016ferroelectric}, and previous calculations report a polarization of $24.9\,\mu\mathrm{C\,cm^{-2}}$ along the $c$ axis~\cite{dong2025nonrelativistic,stoeffler2012first}. The valence and conduction bands show pronounced spin splitting, including at $\Gamma$ (Fig.~\ref{fig:GaFeO3}(b)). Polarization reversal switches the sign of the band-resolved spin polarization without reorienting the ordered Fe moments~\cite{gu2025ferroelectric,dong2025nonrelativistic} (Fig.~\ref{fig:GaFeO3} and Figs.~\textcolor{blue}{S5} and \textcolor{blue}{S6}). The MPG $mm'2'$ also permits linear magnetoelectric coupling.

The projected density of states (DOS) shows predominantly O-$p$ and Fe-$d$ character at the valence- and conduction-band edges, respectively (Fig.~\textcolor{blue}{S7}). Without SOC, conventional antiferromagnets (AFMs) and altermagnets (ALMs) have identical spin-resolved DOS, $D_{\uparrow}(E)=D_{\downarrow}(E)$, whereas CFIM GaFeO$_3$ exhibits $D_{\uparrow}(E)\neq D_{\downarrow}(E)$, reflecting the absence of symmetries interchanging its opposite-spin sublattices. This spin asymmetry underlies the spin-dependent shift and injection photocurrents discussed below.

\begin{figure}[!htp]
\centering
\includegraphics[width=1.0\linewidth]{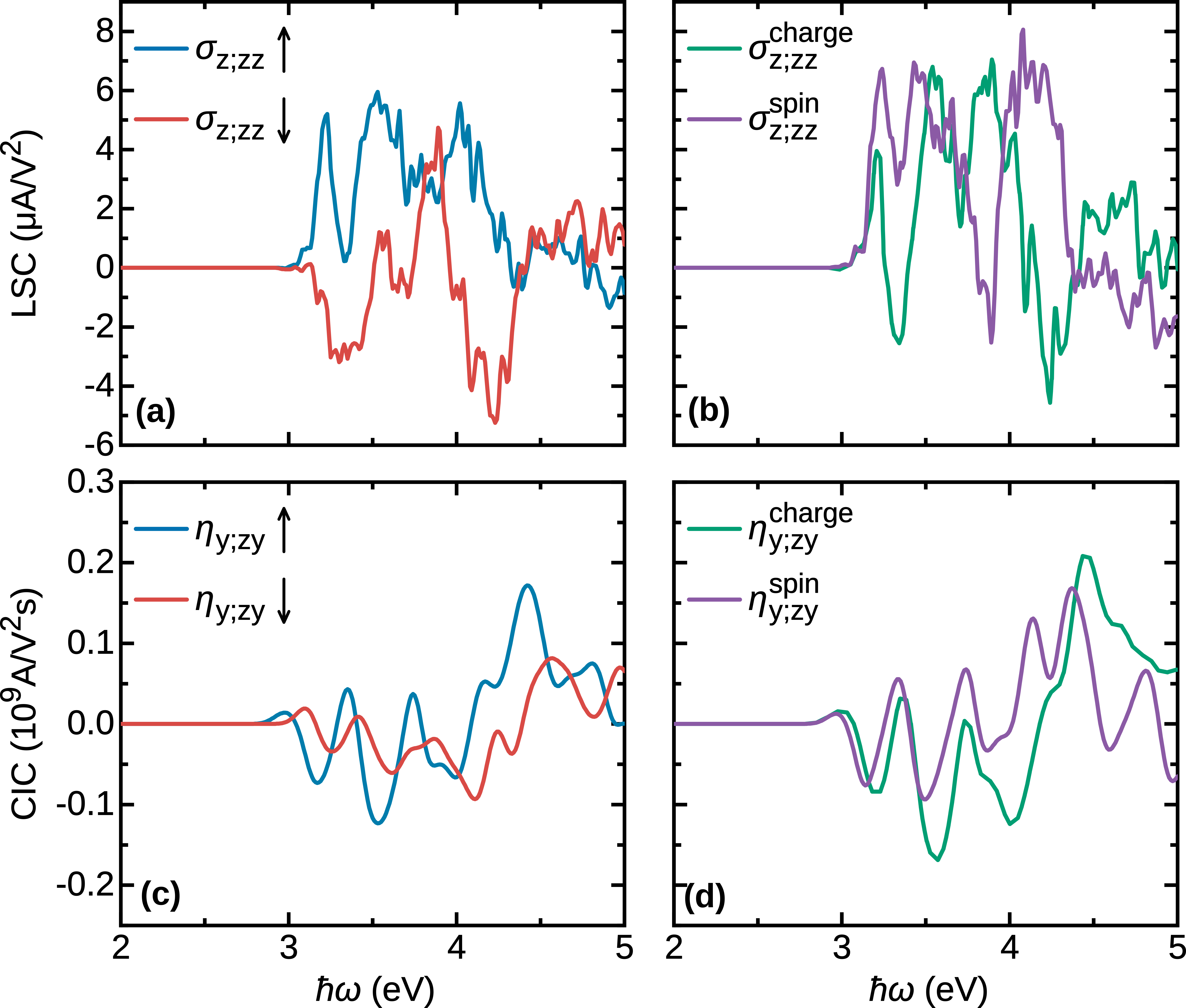}
\caption{Nonlinear photocurrents of GaFeO$_3$ without SOC.
(a,c) Spin-resolved LSC $\sigma_{zzz}$ and CIC $\eta_{yzy}$ spectra, respectively.
(b,d) Corresponding charge and spin responses, obtained from the sum and difference of the spin-channel contributions, respectively.}
\label{fig:spinphotocurrent}
\end{figure}

\underline{\it Spin and Charge Photocurrent Symmetry Analysis--}
Without SOC, GaFeO$_3$ has the oriented SPG ${}^{1}m\,{}^{1}m\,{}^{1}2\times{}^{\infty_{100}m}1$, with ordered moments collinear along $x$. Its spin-only subgroup $\infty_{100}m$ contains arbitrary rotations about $x$ and spin-space mirrors $m_y^{\mathrm{s}}$ and $m_z^{\mathrm{s}}$, normal to $y$ and $z$, respectively. The spin rotation $C_{2x}$ leaves $s_x$ invariant but reverses $s_y$ and $s_z$, without changing the current or optical-field indices, so all $s=y$ and $s=z$ spin-photocurrent components vanish and only spin polarization along the ordered moments ($x$) is allowed (Table~\ref{tab:gafeo3_photocurrent}).

The four spatial operations $R\in\{E,m_x,m_y,2_z\}$ of the SPG enter as $[1\thinspace\Vert\thinspace R]$, with the identity spin part, so the $x$-polarized spin tensors $\sigma^x_{i;jk}$ and $\eta^x_{i;jk}$ transform exactly like their charge counterparts (Tables~\textcolor{blue}{S2--S5} and \textcolor{blue}{S10--S13}). As a result, the spin and charge responses without SOC share the same allowed tensor components within each channel (Table~\ref{tab:gafeo3_photocurrent}). The linear SC and circular IC are allowed in both. The spin-space mirrors $m_y^{\mathrm{s}}$ and $m_z^{\mathrm{s}}$ preserve $s_x$ and have the form $[U_s\thinspace\Vert\thinspace E]$ with $\det U_s=-1$. 
These mirrors act as an effective time reversal within each spin channel, forcing the $\mathcal{T}$-odd LIC and CSC to vanish for both charge and spin. Because the charge and spin responses share the same symmetry-allowed tensor components, the response is generically a spin-polarized charge photocurrent rather than a pure spin current.
This holds for any collinear magnet without a spin-channel-exchanging operation, and CFIMs realize it at zero net magnetization.

\begin{figure}[!htp]
\centering
\includegraphics[width=1.0\linewidth]{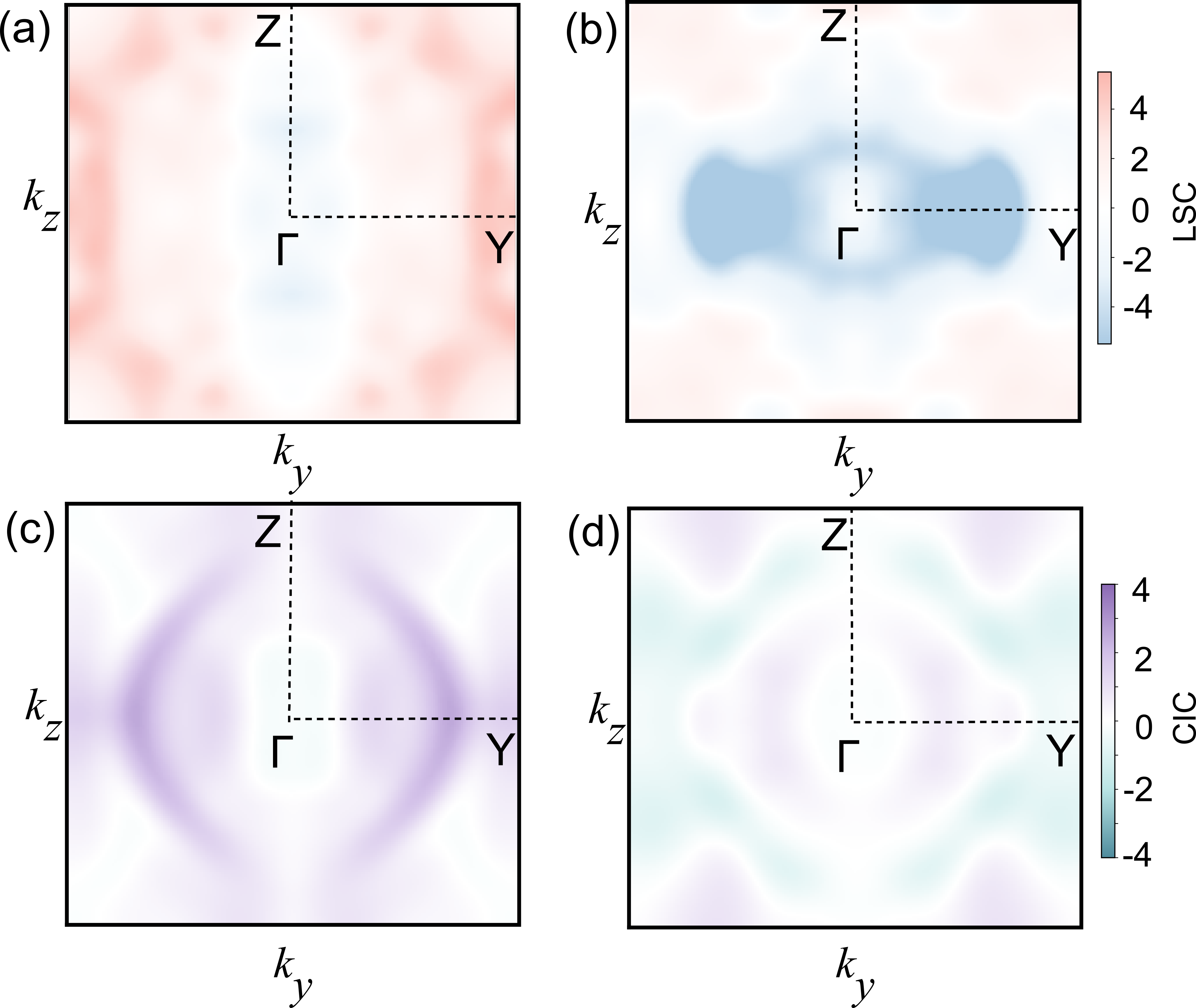}
\caption{Momentum-resolved photocurrent contributions in GaFeO$_3$ without SOC.
(a,b) LSC $\sigma_{z;zz}$ at $\hbar\omega=4.14\,\mathrm{eV}$ and (c,d) CIC $\eta_{y;zy}$ at $4.37\,\mathrm{eV}$ in the $k_y$--$k_z$ plane, integrated over $k_x$.
 Left and right panels show spin-up and spin-down contributions, respectively.}
\label{fig:kresolved}
\end{figure}

SOC couples spin to the lattice, and the charge and spin photocurrent tensors transform under the MPG $mm'2'$ according to their Jahn symbols. For example, the charge and spin LSC tensors have Jahn symbols $V[V^2]$ and $aeVV[V^2]$, respectively. The common factor $V[V^2]$ describes the current and optical-field indices, while the additional factor $aeV$ accounts for spin. For spin polarization along $x$, the signs associated with $a$, $e$, and the spin index cancel under $m_x$, $m_y'$, and $2_z'$, giving $+1$ in every case. SOC also allows LIC and CSC, so all four channels can support photocurrents with spin polarization along $x$, accompanied by charge flow (Table~\ref{tab:gafeo3_photocurrent}). 

For charge, the LSC and CIC components are the same as without SOC, because these $\mathcal{T}$-even tensors see only the spatial operations of $mm2$, whereas the prefix $a$ now permits the magnetic LIC and CSC. For light polarized along a principal axis, the LSC flows along the polar axis $z$ and the LIC along $y$, perpendicular to both the polarization and the moments, so the two are separated by direction.
This direction is that of the toroidal moment, the only $\mathcal{T}$-odd polar vector allowed by $mm'2'$. For light propagating along it, nonreciprocal absorption, the optical magnetoelectric effect, has been observed in GaFeO$_3$~\cite{jung2004optical}.
For circularly polarized light in the $xy$ plane, the helicity-dependent current is purely CSC, $\sigma_{x;xy}$.

\begin{figure}[!htp]
\centering
\includegraphics[width=1.0\linewidth]{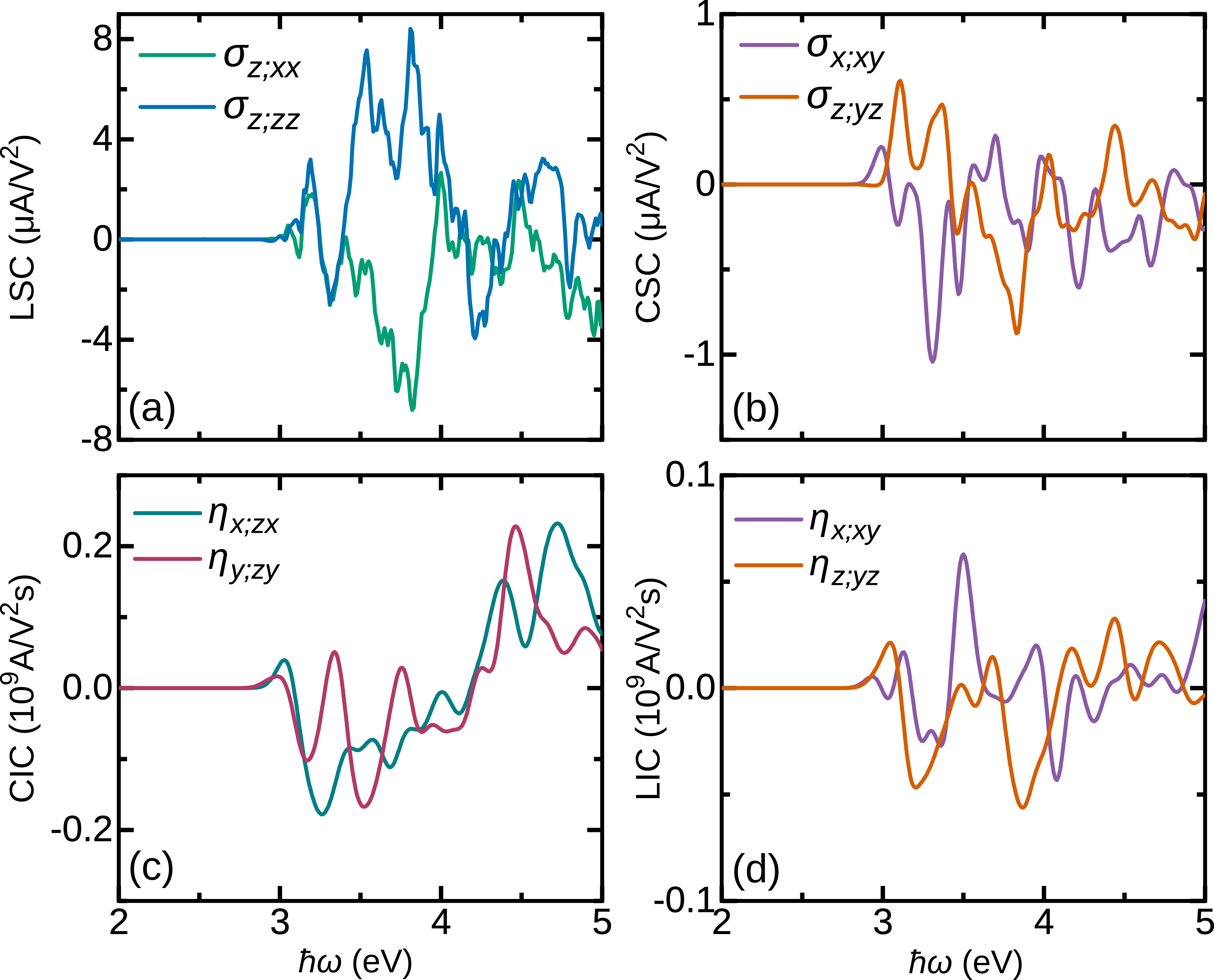}
\caption{Charge photocurrent spectra of GaFeO$_3$ with SOC.
(a) LSC, (b) CSC, (c) CIC, and (d) LIC. The remaining components are provided in the SI.}
\label{fig:chargephotocurrent}
\end{figure}

In contrast, the additional $aeV$ factor gives $(-,-,+)$ and $(-,+,-)$ for spin polarization along $y$ and $z$ under $(m_x,m_y',2_z')$, respectively. These spin polarizations are perpendicular to the ordered moments along $x$, while the current direction is specified independently by the index $i$. Under $m_x$, charge and $s_x$ components require an even number of $x$ among $i,j,k$, whereas $s_y$ and $s_z$ components require an odd number, so components supporting these spin polarizations are forbidden for the charge and $x$-polarized spin responses in all four channels (Table~\ref{tab:gafeo3_photocurrent}). For example, for light polarized within the $yz$ plane ($E_x=0$), every charge and $s_x$ component with current along $x$ vanishes in all four channels, whereas components such as $\sigma^z_{x;yy}$ (LSC), $\eta^y_{x;yy}$ (LIC), $\eta^y_{x;yz}$ (CIC), and $\sigma^z_{x;yz}$ (CSC) remain allowed. The photocurrent along $x$ is then a charge-free pure spin current polarized perpendicular to the ordered moments, which is absent without SOC. Spin photocurrents with SOC are discussed based solely on symmetry analysis, with their explicit calculation beyond the scope of this work.

\underline{\it Charge and Spin Photocurrents without SOC--}
We now quantify the charge and spin responses allowed by the SPG without SOC. Figure~\ref{fig:spinphotocurrent} shows representative LSC $\sigma_{z,zz}$ and CIC $\eta_{y,zy}$ spectra. For both components, the spin-up and spin-down contributions exhibit distinct amplitudes and spectral profiles (Fig.~\ref{fig:spinphotocurrent}(a,c)). Their imbalance produces spin responses comparable in magnitude to their charge counterparts, with the spin LSC reaching approximately $8\,\mu\mathrm{A\,V^{-2}}$ near $\hbar\omega=4.1\,\mathrm{eV}$ (Fig.~\ref{fig:spinphotocurrent}(b,d)). Where the two spin-channel contributions have opposite signs, they reinforce the spin response while partially cancelling in the charge response; where they have the same sign, as in the CIC, their unequal magnitudes still leave a finite spin response. Magnetic compensation is thus compatible with strong optical spin selectivity and substantial nonlinear spin transport without SOC. Both spin responses are polarized along the ordered Fe moments and coexist with charge flow, consistent with the SPG analysis.

\begin{figure}[!htbp]
\centering
\includegraphics[width=1.0\linewidth]{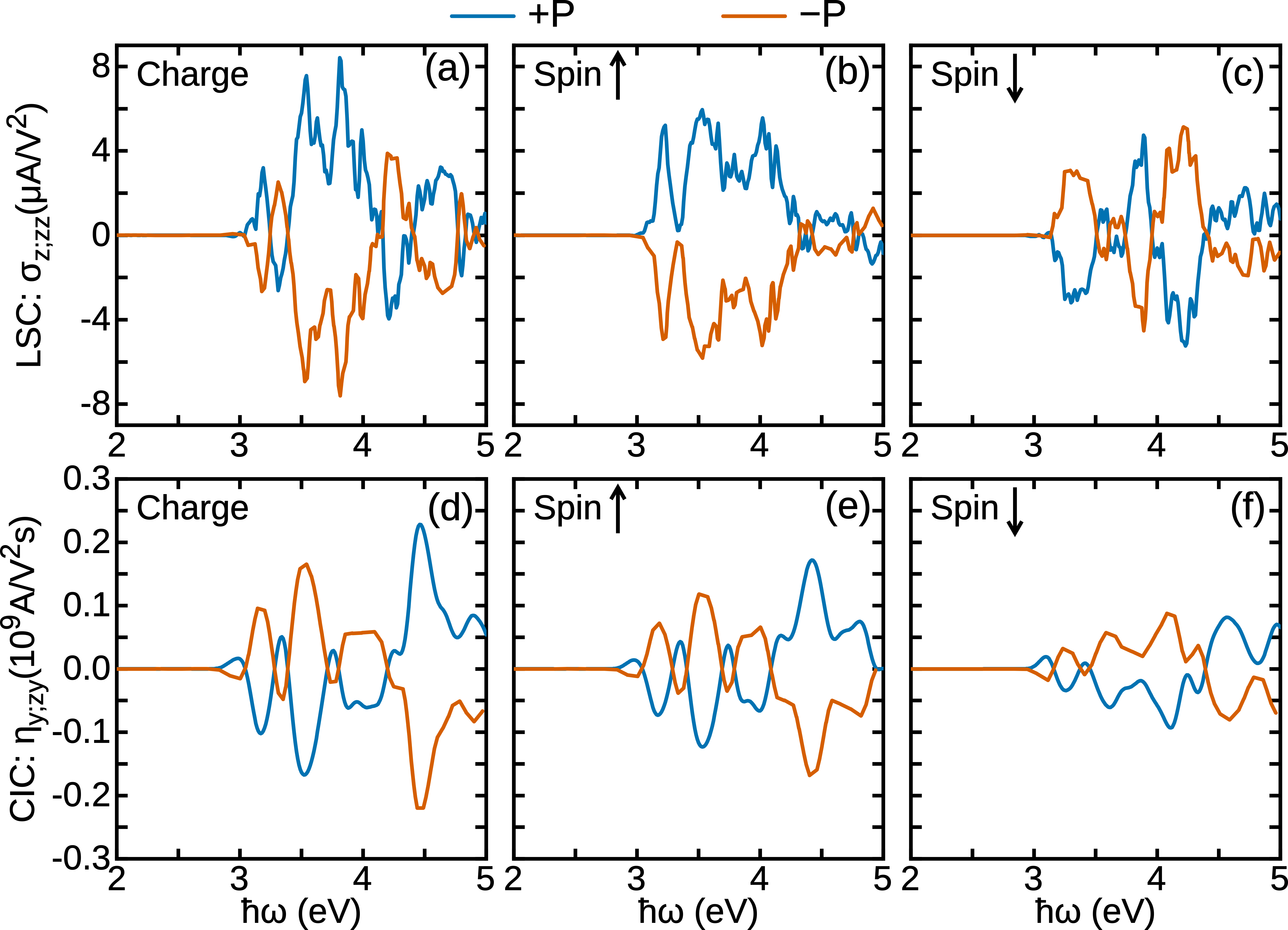}
\caption{Ferroelectric control of charge and spin-resolved photocurrents in GaFeO$_3$ without SOC for the $+P$ and $-P$ states. (a--c) LSC $\sigma_{z,zz}$ and (d--f) CIC $\eta_{y,zy}$. The columns show charge, spin-up, and spin-down responses, respectively.
The remaining components are provided in the SI.}
\label{fig:FEphotocurrent}
\end{figure}

To identify the momentum regions underlying this spin-channel imbalance, we examine the momentum-resolved LSC and CIC contributions in the $k_y$--$k_z$ plane, integrated over $k_x$, without SOC (Fig.~\ref{fig:kresolved}). At $\hbar\omega=4.14\,\mathrm{eV}$, close to the spin LSC maximum, $\sigma_{z,zz}^{\uparrow}$ is broadly positive away from $\Gamma$, whereas $\sigma_{z,zz}^{\downarrow}$ is dominated by negative lobes extending along $k_y$ around $\Gamma$ (Fig.~\ref{fig:kresolved}(a,b)). These positive spin-up and negative spin-down weights reinforce the spin-channel difference. Spin selectivity is also evident in the CIC at $\hbar\omega=4.37\,\mathrm{eV}$. Here, $\eta_{y,zy}^{\uparrow}$ exhibits extended positive lobes, while $\eta_{y,zy}^{\downarrow}$ contains positive and negative contributions distributed across different momentum regions (Fig.~\ref{fig:kresolved}(c,d)). Together, these distributions provide a microscopic picture of how unequal momentum-dependent contributions from the two spin channels produce the finite spin photocurrents.

\underline{\it Charge Photocurrents with SOC--}
We next quantify the charge photocurrents of GaFeO$_3$ with SOC in all four channels, as summarized in Fig.~\ref{fig:chargephotocurrent}. The LSC components $\sigma_{z,xx}$ and $\sigma_{z,zz}$ exhibit pronounced spectral features, with $\sigma_{z,zz}$ reaching approximately
$8.5\,\mu\mathrm{A\,V^{-2}}$ at a photon energy of $\hbar\omega \approx 3.82\,\mathrm{eV}$
(Fig.~\ref{fig:chargephotocurrent}(a)). The CSC response is comparatively weaker, with the displayed components reaching approximately $1\,\mu\mathrm{A\,V^{-2}}$ while remaining finite over a broad photon-energy range (Fig.~\ref{fig:chargephotocurrent}(b)). The dependence on light polarization evident in the shift currents also extends to the injection-current channels. The displayed CIC coefficients reach approximately $0.2\times10^{9}\,\mathrm{A\,V^{-2}\,s^{-1}}$, corresponding to about $0.2\,\mu\mathrm{A\,V^{-2}}$ for $\tau=1\,\mathrm{fs}$, and are larger in magnitude than the LIC coefficients (Fig.~\ref{fig:chargephotocurrent}(c,d)). These channels also differ in their symmetry requirements. LSC and CIC are allowed in noncentrosymmetric crystals without SOC and do not require magnetic order. In contrast, CSC and LIC are magnetic responses. In GaFeO$_3$, SOC enables these additional channels, consistent with our MPG analysis, while leaving the LSC and CIC nearly unchanged (Fig.~\ref{fig:spinphotocurrent}(b,d)).

\underline{\it Ferroelectric Photocurrent Switching--}
We finally examine how ferroelectric switching controls the charge and spin photocurrents in GaFeO$_3$. Reversing the ferroelectric polarization from $+P$ to $-P$ reverses the sign of all charge LSC and CIC tensor components, as illustrated by the representative $\sigma_{z,zz}$ and $\eta_{y,zy}$ spectra (Fig.~\ref{fig:FEphotocurrent}(a,d)). Without SOC, the spin-up and spin-down contributions are interchanged with reversed sign, $\sigma_{\uparrow}(-P)=-\sigma_{\downarrow}(+P)$, for both LSC (Fig.~\ref{fig:FEphotocurrent}(b,c)) and CIC (Fig.~\ref{fig:FEphotocurrent}(e,f)). Ferroelectric switching therefore reverses the charge current and its spin polarization, while the spin response $\sigma_{\uparrow}-\sigma_{\downarrow}$ keeps its sign. This behavior is consistent with the polarization dependence of the spin-resolved electronic structure discussed above and occurs while the ordered Fe moments retain their orientations. Such switchable photocurrents could support electrically programmable photodetection and nonvolatile memory in which the ferroelectric state is read through the photovoltaic response~\cite{gong2025reconfigurable,guo2013non}.

\underline{\it Conclusions--}
We have shown that the spin group and the magnetic group impose different selection rules on the nonlinear charge and spin photocurrents of the polar compensated ferrimagnet GaFeO$_3$. Without SOC, only linear shift and circular injection currents are allowed, with identical components for charge and spin. The spin current is polarized along the Fe moments and always accompanies a charge current. Its shift component reaches about $8\,\mu\mathrm{A\,V^{-2}}$, comparable to the charge response, despite zero net magnetization. SOC leaves these charge currents nearly unchanged, with the linear shift current reaching about $8.5\,\mu\mathrm{A\,V^{-2}}$, and activates magnetic charge linear injection and circular shift currents, the latter reaching about $1\,\mu\mathrm{A\,V^{-2}}$. By symmetry alone, the magnetic group further predicts pure spin currents polarized perpendicular to the moments, which the spin group forbids. Ferroelectric switching reverses the linear shift and circular injection charge currents and their spin polarization, but not the spin currents, while the Fe moments remain fixed. A complete description of charge and spin photocurrents in compensated ferrimagnets thus requires the magnetic group. Polar members of this class are candidates for electrically reconfigurable charge and spin photovoltaics.

\underline{\it Acknowledgments--}  
The authors acknowledge the computing time provided on the Lichtenberg high-performance computer by the NHR Center NHR4CES at TU Darmstadt. Fu Li acknowledges support from the China Scholarship Council. Hongbin Zhang acknowledges support from the Deutsche Forschungsgemeinschaft (DFG, German Research Foundation), Project-ID 463184206. Harish K. Singh acknowledges the EuroHPC Joint Undertaking for granting access to the VEGA supercomputer at IZUM, Slovenia, through project EHPC-REG-2025R01-197-EuroHPC.

\bibliography{reference}

\end{document}